\documentclass[aps,prd,showpacs,onecolumn,preprintnumbers,notitlepage,superscriptaddress,nofootinbib,amsmath,amsfont,amssymb,graphicx,10pt,singlespacing]{revtex4} 
\pdfoutput=1 
\usepackage{amsmath} 
\usepackage{bm}
\usepackage{times} 
\usepackage{braket} 
\usepackage{color,graphicx}
\usepackage{slashed} 
\usepackage{hyperref}

\newcommand{\qq}{q^2}
\newcommand{\nn}{\nonumber}
\newcommand {\E}[1]{\times 10^{#1}} 
\newcommand {\e}[1]{\mathrm{~#1}} 
\newcommand{\mc}[1]{\mathcal{#1}}

\newcommand{\mrm}[1]{\mathrm{#1}}
\newcommand{\trm}[1]{\textrm{#1}}

\renewcommand{\Re}[0]{\mrm{Re}}
\renewcommand{\Im}[0]{\mrm{Im}}

\definecolor{Red}{rgb}{1.,0.,0.}

\definecolor{Blue}{rgb}{0.,0.,1.}

\definecolor{nicered}{rgb}{0.7,0.1,0.1}
\definecolor{nicegreen}{rgb}{0.1,0.5,0.1}

\hypersetup{colorlinks,citecolor=nicegreen,linkcolor=nicered}

\begin{document}
\title{Resonance catalyzed CP asymmetries in $D \to P \ell^+ \ell^-$}

\author{Svjetlana Fajfer} \email[Electronic
address:]{svjetlana.fajfer@ijs.si} 
\affiliation{Department of Physics,
  University of Ljubljana, Jadranska 19, 1000 Ljubljana, Slovenia}
\affiliation{J. Stefan Institute, Jamova 39, P. O. Box 3000, 1001
  Ljubljana, Slovenia}

\author{Nejc Ko\v snik} 
\email[Electronic address:]{nejc.kosnik@ijs.si}
\affiliation{Laboratoire de l'Acc\'el\'erateur Lin\'eaire,
Centre d'Orsay, Universit\'e de Paris-Sud XI,
B.P. 34, B\^atiment 200,
91898 Orsay cedex, France}
\affiliation{J. Stefan Institute, Jamova 39, P. O. Box 3000, 1001 Ljubljana, Slovenia}

\begin{abstract}
  Recently observed increase of direct CP asymmetry in charm meson
  nonleptonic decays is difficult to explain within the SM. If this
  effect is induced by new physics, this might be investigated in
  other charm processes. We propose to investigate new CP violating
  effects in rare decays $D \to P \ell^+ \ell^-$, which arise due to
  the interference of resonant part of the long distance contribution
  and the new physics affected short distance contribution. Performing
  a model independent analysis, we identify as appropriate observables
  the differential direct CP asymmetry and partial decay width CP
  asymmetry. We find that in the most promising decays $D^+ \to \pi^+
  \ell^+ \ell^-$ and $D_s^+ \to K^+ \ell^+ \ell^-$ the
  ``peak-symmetric'' and ``peak-antisymmetric'' CP asymmetries are strong
  phase dependent and can be of the order $1\,\%$ and $10\,\%$, respectively.
\end{abstract}

\preprint{LAL 12-275}
\maketitle

\section{Introduction}
In last two decades chances to observe new physics in charm processes
were considered to be very small. In the case of flavor changing
neutral current processes the Glashow-Iliopoulos-Maiani~(GIM)
mechanism plays a significant role, leading to cancellations of
contributions of $s$ and $d$ quarks, while intermediate $b$ quark
contribution is suppressed by $V_{ub}$ element of the
Cabibbo-Kobayashi-Maskawa~(CKM) matrix. However, this has changed at
the end of last year when LHCb experiment reported a non-vanishing
direct CP asymmetry in $D^0 \to K^+ K^-$ and $D^0 \to \pi^+
\pi^-$~\cite{Aaij:2011in} also confirmed by the CDF
experiment~\cite{Collaboration:2012qw}. The lack of appropriate
theoretical tools to handle long distance dynamics in these processes
is even more pronounced than in the case of $B$ mesons due to
abundance of charmless resonances with the masses close to the masses
of charm mesons. Many papers investigated whether this result can be
accommodated within the standard model (SM) or is it new physics~(NP)
that causes such an effect. The measured difference between the CP
asymmetry in $D^0 \to K^+ K^-$ and $D^0 \to \pi^+ \pi^-$ is a factor
$5-10$ larger than expected in the SM and eventually can be a result
of nonperturbative QCD dynamics as pointed out in
refs.~\cite{Golden:1989qx,Brod:2011re,Pirtskhalava:2011va,Bhattacharya:2012ah,Feldmann:2012js,Brod:2012ud,Cheng:2012wr,Cheng:2012xb}.
Model independent studies~\cite{Isidori:2011qw,Gedalia:2012pi}
indicated that among operators describing NP effect, the most likely
candidate is the effective $\Delta C = 1$ chromomagnetic dipole
operator. In order to distinguish between SM or NP scenarios as
explanation of the observed phenomena it is crucial to investigate
experimentally and theoretically all possible processes in which the
same operator might contribute. Recently the effects of the same kind
of new physics have been explored in radiative~\cite{Isidori:2012yx}
and inclusive charm decays with a lepton pair in the final
state~\cite{Paul:2011ar}. In~\cite{Isidori:2012yx} it was found that
NP induces an enhancement of the matrix elements of the
electromagnetic dipole operators leading to CP asymmetries of the
order of few percent.

In addition to radiative weak decays, charm meson decays to a light
meson and leptonic pair might serve as a testing ground for CP
violating new physics contributions. As in other weak decays of charm
mesons the long distance dynamics dominates the decay widths of $D \to
P \ell^+ \ell^-$~\cite{Burdman:2001tf,Fajfer:2005ke,Fajfer:2007dy} and
it requires special task to find the appropriate variables containing
mainly short-distance contributions.  In this study we investigate
partial decay width CP asymmetry in the case of $D \to P \ell^+
\ell^-$ decay.  The short distance dynamics is described by effective
operators $\mc{O}_7$, $\mc{O}_9$, and $\mc{O}_{10}$ of which the
electromagnetic dipole operator $\mc{O}_7$ carries a CP odd phase of
beyond the SM origin, developed due to mixing under QCD
renormalization with the chromomagnetic operator. In this paper we
investigate impact of this mixing on the $D \to P \ell^+ \ell^-$ decay
dynamics. The paper is organized as follows: Section~\ref{sec-SD}
contains the description of the short distance contributions and
hadronic form factors, Sec.~\ref{sec-LD} is devoted to the long
distance dynamics. In Sec.~\ref{sec-asymm} we present the partial
width asymmetry. We summarize our findings in Sec.~\ref{sec-summ}.

\section{Effective Hamiltonian and Short Distance Amplitude}
\label{sec-SD}
The 
dynamics of $c \to u \ell^+ \ell^-$ decay on scale $\sim m_c$ is
defined by the effective
Hamiltonian~\cite{Burdman:2001tf,Isidori:2011qw}
\begin{align}
  \label{eq:Heff}
  \mc{H}_\mrm{eff} &= \lambda_d \mc{H}^d + \lambda_s \mc{H}^s + \lambda_b \mc{H}^\mrm{peng}\,,
\end{align}
where the CKM weights are $\lambda_i = V_{ci}^* V_{ui}$. For the first two
generations we have the current-current operators
\begin{align}
  \label{eq:2}
  \mc{H}^{q=d,s} &= -\frac{4 G_F}{\sqrt{2}} (C_1 \mc{O}_1^q + C_2
  \mc{O}_2^q)\,,\\
  \mc{O}_1^q &= (\bar q^\alpha_L \gamma^\mu c^\alpha_L)\,(\bar u^\beta_L \gamma_\mu
  q^\beta_L)\,,\nn\\
  \mc{O}_2^q &= (\bar q^\alpha_L \gamma^\mu c^\beta_L)\,(\bar u^\beta_L \gamma_\mu q^\alpha_L)\nn\,,
\end{align}
with color indices $\alpha$, $\beta$. The effects of $b$ quark and
heavier particles are contained within
the set operators of dimension-5 and 6
\begin{equation}
  \label{eq:3}
  \mc{H}^\mrm{peng} = -\frac{4 G_F}{\sqrt{2}} \sum_{i=3,\ldots,10} C_i \mc{O}_i\,,
\end{equation}
where electromagnetic (chromomagnetic) penguins and electroweak
penguins/boxes with leptons are
\begin{align}
  \mc{O}_7 &= \frac{e m_c}{(4\pi)^2}\, \bar u \sigma_{\mu\nu} P_R c
  \,F^{\mu\nu}\,,\\
  \mc{O}_8 &= \frac{g m_c}{(4\pi)^2}\, \bar u \sigma_{\mu\nu} P_R c
  \,G^{\mu\nu}\nn\,,\\
  \mc{O}_9 &= \frac{e^2}{(4\pi)^2}\, (\bar u \gamma^\mu P_L c)
  (\bar\ell \gamma_\mu \ell)\nn\,,\\
  \mc{O}_{10} &= \frac{e^2}{(4\pi)^2}\, (\bar u \gamma^\mu P_L c)
  (\bar\ell \gamma_\mu \gamma_5 \ell)\nn\,.
\end{align}
Complete set of QCD penguin operators $\mc{O}_{3,...,6}$ can be found
in refs.~\cite{Burdman:1995te,Burdman:2001tf}. Decay width spectrum of
$c \to u \ell^+ \ell^-$ is dominated by the two light generations'
effective Hamiltonians, $\mc{H}^{d,s}$, and is exactly CP-even when
$\lambda_d + \lambda_s = 0$ holds. Only when we include the third
generation we get a possibility of having a nonvanishing imaginary
part: $\Im (\lambda_{b}/\lambda_d) = -\Im
(\lambda_{s}/\lambda_d)$. However, the CP violating parts of the
amplitude are suppressed by a tiny factor $\lambda_b/\lambda_d \sim
10^{-3}$ with respect to the CP conserving ones and only tiny effects
of CP violation is expected. On the other hand, too large direct CP
is measured in singly Cabibbo suppressed decays $D^0 \to \pi
\pi,KK$. Should this enhancement be due to new physics, one can most
naturally satisfy other flavor constraints by assigning a NP
contribution to the chromomagnetic operator $\mc{O}_8$ at some high
scale above $m_t$~\cite{Isidori:2011qw}. In this case one must also
get $C_{7}(m_c)$ that carries related new physics CP phase due to
mixing of $\mc{O}_8$ into $\mc{O}_7$ under QCD renormalization.
We shall consider the range proposed in~\cite{Isidori:2012yx},
\begin{equation}
  \label{eq:1}
  |\Im \,\left[\lambda_b C_7(m_c)\right]| = (0.2-0.8)\E{-2}\,,
\end{equation}
where the authors used this particular value to estimate the size of
direct CP violation in $D \to P \gamma$ decays. This approach was
further scrutinized recently in~\cite{Lyon:2012fk}.

We define the short distance amplitude as the one coming from
operators $\mc{O}_{7}$, $\mc{O}_9$, and $\mc{O}_{10}$ (they do not
contain, apart from $c$ and $u$ fields, any colored degrees of
freedom). While their contribution to the decay width is negligible in
the resonance-dominated regions due to small CKM elements, possible
imaginary parts of Wilson coefficients may generate direct CP
violation via interference with the CP-even long distance amplitude
(that we define below). 
In light of the above discussion we will assume that in the SD
amplitude only $\mc{O}_7$ carries a CP-violating phase. Relevant SD amplitude of $D
\to \pi \ell^+ \ell^-$, where $\ell = e, \mu$, is then
\begin{equation}
  \label{eq:SDAmpl}
  \mc{A}_{\mathrm{SD}}^\mrm{CPV} = -\frac{i\sqrt{2}G_F \alpha}{\pi}
  \,\lambda_b C_7(m_c)\, \frac{m_c}{m_D+m_\pi}f_T(q^2)\,\bar u(k_-)
  \slashed{p} v(k_+)\,,
\end{equation}
where $p$ is momentum of the $D$ meson and $q = k_- + k_+$ is
momentum of the lepton pair. The form factors for $D \to \pi$
transition via vector current and electromagnetic dipole operators are
defined as customary
\begin{align}
  \label{eq:FF}
\langle \pi(p^\prime)| \bar{u} \gamma_{\mu} c |D(p)\rangle & =  \left[ (p + p^\prime)_\mu - \frac{m_D^2 - m_\pi^2}{q^2} q_\mu \right] F_{1}(q^2) + \frac{m_D^2-m_\pi^2}{q^2} q_{\mu} F_{0}(q^2) \, , \nn \\
  \langle \pi(p^\prime)| \bar{u}\sigma_{\mu\nu} c |D(p) \rangle &= -i \left( p_\mu p^\prime_\nu  - p_\nu p^\prime_\mu \right) \frac{2 f_T(q^2)}{m_D + m_\pi} \,,
\end{align}
with $q^2 = (p-p^\prime)^2$. 

\subsection{Parameterization of the tensor form factor}
The lattice QCD calculations of the form factors for the semileptonic
$D \to \pi$ transitions are rather well known (see
e.g.~\cite{Koponen:2012di}) and their analysis are based on the use of
$z$-parametrization~\cite{Becher:2005bg,Bourrely:2008za}. The
$z$-parametrization of the $D \to P$ form factors in practical
use is often replaced by the Be\v cirevi\'c-Kaidalov~(BK)
parametrization~\cite{Becirevic:1999kt} (as in~\cite{Aubert:2007wg}
and~\cite{Widhalm:2006wz}). 
Quenched lattice QCD results exist for $F_{1,0}$ as well for the tensor
form factor~\cite{Abada:2000ty,DB} and are presented in the BK
parameterization:
\begin{align}
  \label{eq:bk}
  F_1(q^2) &= \frac{F_1(0)}{\left(1-\frac{q^2}{m_{D^*}^2}\right)
    \left(1-a \frac{q^2}{m_{D^*}^2}\right)}\,,\\
  F_0(q^2) &= \frac{F_1(0)}{1-\frac{1}{b}\frac{q^2}{m_{D^*}^2}}\,,\nn\\\nn\\
  \label{eq:bkpars}
  F_1(0) &=  0.57(6)\,,\\
  a &= 0.18(17)\,,\nn\\
  b &= 1.27(17)\,.\nn
\end{align}
For $f_T(q^2; \mu)$ it has recently been noted that in the high $q^2$
region the $B \to K$ matrix elements are well described by the nearest
pole ansatz for form factors $F_1$ and $f_T$ (see Appendix A
of~\cite{Becirevic:2012fy}). Analogously we expect a dominance of the
$D^*$ resonance for $F_1(q^2)$ and $f_T(q^2;\mu)$ close to the
zero-recoil point and consequently the ratio of the two form factors
becomes a constant.
The following scale invariant function
\begin{equation}
  \label{eq:fTtilde}
  \tilde f_T(q^2) \equiv \frac{m_{D^*}}{m_D + m_\pi} \frac{f_{D^*}^V}{f_{D^*}^T (\mu)}\, f_T(q^2; \mu)\,,
\end{equation}
approaches $F_1(q^2)$ at large $q^2$. Here $f_{D^*}^V$ and
$f_{D^*}^T(\mu)$ are the decay constants of $D^*$ via the vector and
tensor currents, respectively. A fit of the lattice
data~\cite{Abada:2000ty,DB} to the BK shape
\begin{align}
  \label{eq:fTtildeBK}
  \tilde f_T(q^2) &= \frac{\tilde f_T(0)}{\left(1-\frac{q^2}{m_{D^*}^2}\right)
    \left(1-a_T \frac{q^2}{m_{D^*}^2}\right)}\,, \\
  \tilde f_T (0) &= 0.56(5)\,, \qquad a_T = 0.18(16)\,,\nn
\end{align}
tells us that within the errors the form factor is single
pole-like. Extrapolation to the low $q^2$ region, which is more relevant for
our discussion, gives $\left.f_T(q^2;\mu)/F_1(q^2)\right|_{q^2=0} =
0.83 \pm 0.19$ that is marginally compatible with the results expected
in the Large Energy Effective Theory
limit, where one expects the same ratio to be $1+m_\pi/m_D = 1.07$~\cite{Charles:1998dr,Ball:2004ye}.
The ratio of tensor and vector decay constants,
needed in formula~\eqref{eq:fTtilde} at the charm scale, is
\begin{equation}
  \frac{f_{D^*}^T(\mu=2\e{GeV})}{f_{D^*}^V} = 0.82(3)\,.
\end{equation}

\section{Long distance amplitude}
\label{sec-LD}
Close to the $\phi$ resonant peak the long distance amplitude is, to a
good approximation, driven by nonfactorizable contributions of
four-quark operators in $\mc{H}^s$. The width of $\phi$ resonance is
very narrow ($\Gamma_\phi/m_\phi \approx 4\E{-3}$) and well separated
from other vector resonances in the $\qq$ spectrum of $D \to P \ell^+
\ell^-$. Relying on vector meson dominance hypothesis the
$q^2$-dependence of the decay spectrum close to the resonant peak
follows the Breit-Wigner
shape~\cite{Burdman:2001tf,Fajfer:2005ke,Fajfer:2007dy}
\begin{equation}
\label{eq:BW}
  \mc{A}_{\mathrm{LD}}^\phi \left[D  \to \pi  \phi \to  \pi \ell^- \ell^+\right]
   = \frac{i G_F }{\sqrt{2}} \lambda_s \frac{8\pi \alpha}{3}\, a_{\phi}  e^{i \delta_{\phi}}\,
   \frac{ m_{\phi} \Gamma_\phi   }{q^2-m_{\phi}^2 + i m_{\phi} \Gamma_{\phi}}\, \bar{u}(k_-)\, \slashed{p}\,v(k_+)\,.
\end{equation}
Here we use $\alpha = 1/137$ in the leading order in electromagnetic
interaction.

The long distance amplitude is also affected by nonfactorizable effects of
four-quark operators $\mc{O}_{3-6}$ and by the gluonic penguin operator
$\mc{O}_8$.
 %
Whereas the former have only tiny CP violation and are suppressed with
$\lambda_b/\lambda_s $ compared to~\eqref{eq:BW}, the $\mc{O}_8$ contribution
can be important for the results of this study since NP CP-odd phases
present in Wilson coefficients $C_7$ and $C_8$ are closely
related. Opposed to the $\mc{O}_7$ mediated amplitude with a single
photon exchange the $\mc{O}_8$ amplitude necessarily involves a strong
loop suppression factor of the order $\alpha_s(\mu = m_c)/\pi$ and is
therefore subdominant in this perturbative picture.  However, in the
full nonperturbative treatment we cannot exclude an order of magnitude
enhancement of amplitude with $\mc{O}_8$ insertion\footnote{Analogous
  nonfactorizable amplitudes of $\mc{O}_8$ in B physics have been
  studied in the framework of QCD
  factorization~\cite{Beneke:1999br,Beneke:2000ry} in $B \to V^{*}
  \ell^+ \ell^-$ decay
  modes~\cite{Beneke:2001at,Beneke:2004dp,Altmannshofer:2008dz}.}. In
this work we will neglect such contributions and therefore our
conclusions will be quantitatively valid provided there is no
nonperturbative enhancement of the $\mc{O}_8$ amplitude.

Finite width of the resonance generates a $q^2$-dependent strong phase
that varies across the peak. We have also introduced the strong phase
on peak, $\delta_\phi$, and the normalization, $a_\phi$, that are both
assumed to be independent of $q^2$. Parameter $a_\phi$ is real and can
be fixed from measured branching fractions of $D \to \pi \phi$ and
$\phi \to \ell^+ \ell^-$ decays~\cite{Fajfer:2007dy}. For definiteness
we will focus on the $\ell = \mu$ decay modes. From the Particle Data
Group compilation we read~\cite{Nakamura:2010zzi}
\begin{align}
 \mrm{Br}(D^+ \to \phi \pi^+) &= (2.65 \pm 0.09 )\times 10^{-3}\,,\\
 \mrm{Br}(\phi \to \mu^+ \mu^-) &= (0.287 \pm 0.019)\times 10^{-3}\,,\nn
\end{align}
and when we take into account the small width of $\phi$
\begin{equation}
   \mrm{Br}(D^+ \to \pi^+ \phi( \to \mu^+ \mu^-)) \approx \mrm{Br}(D^+ \to \phi \pi^+) \times \mrm{Br}(\phi \to \mu^+ \mu^-)\,,
\end{equation}
we find from eq.~\eqref{eq:BW} 
\begin{equation}
  \label{eq:aphi}
a_\phi = 1.23\pm 0.05\,.  
\end{equation}

\section{Direct CP asymmetry}
\label{sec-asymm}
\begin{figure}[!!hct]
  \centering
  \begin{tabular}{r}
  \includegraphics[width=0.6\textwidth]{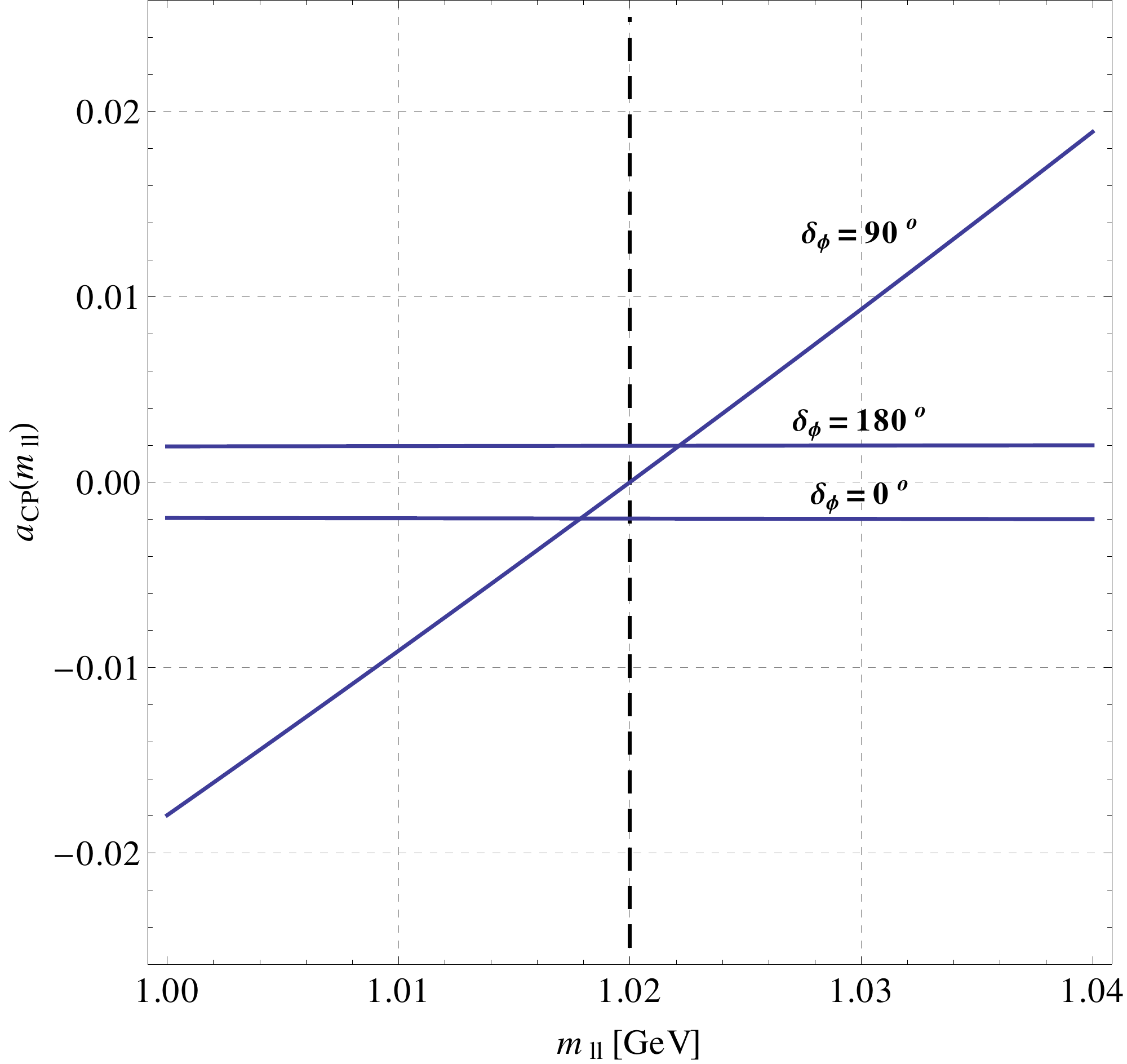}\\    \\
  \includegraphics[width=0.678\textwidth]{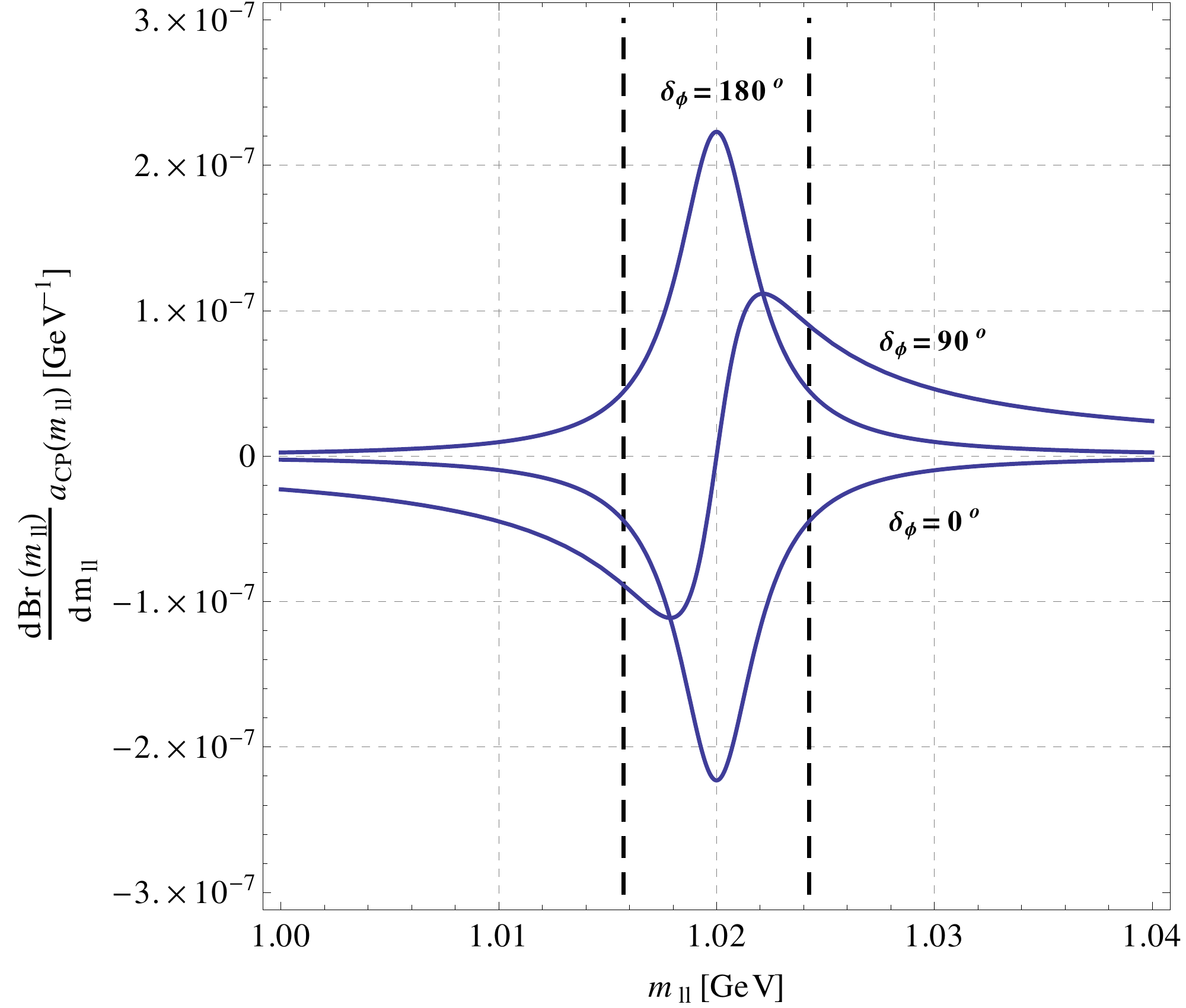}
  \end{tabular}
  \caption{Top: CP asymmetry $a_\mrm{CP} (m_{\ell\ell})$ around the
    $\phi$ resonance (dashed vertical line) for representative values of strong phase
    $\delta_\phi = 0,\pi/2,\pi$. Bottom:
    $(d\mrm{Br}/dm_{\ell\ell})\,a_\mrm{CP} (m_{\ell\ell})$, the
    measure of sensitivity to direct CP-violation. Dashed vertical
    lines at $m_{\ell\ell} = m_\phi \pm \Gamma_\phi$ denote the width
    of the resonance.}
  \label{fig:aCPdiff}
\end{figure}

The direct CP violation in the resonant region is driven by the
interference between the CP-odd imaginary part of the SD amplitude and
the LD amplitude. The pair of CP-conjugated amplitudes read
 \begin{align}
  \mc{A}(D^+ \to \pi^+ \ell^+ \ell^-) &= \mc{A}^\phi_\mrm{LD} + \mc{A}_\mrm{SD}^\mrm{CPV}\,,\\
  \overline{\mc{A}}(D^- \to \pi^- \ell^+ \ell^-) &= \mc{A}^\phi_\mrm{LD} + \overline{\mc{A}}_\mrm{SD}^\mrm{CPV}\,,\nn
 \end{align}
 In principle the short-distance amplitude contains a strong phase
 that can be rotated away because the overall phase of the total
 amplitude is irrelevant. The CP-odd part of the LD amplitude is
 proportional to the imaginary part of the relevant CKM factor $\lambda_s$
that can be safely neglected and accordingly we have put $\mc{A}_{LD}^\phi
= \bar{\mc{A}}_{LD}^\phi$. Then the differential direct CP violation reads
\begin{align}
  \label{eq:acp}
  a_{CP} (\sqrt{q^2}) &\equiv
  \frac{|\mc{A}|^2-|\overline{\mc{A}}|^2}{|\mc{A}|^2+|\overline{\mc{A}}|^2}\\
 &= \frac{-3}{2\pi^2}   \frac{f_T(q^2)}{a_\phi} \frac{m_c}{m_D+m_\pi} 
\Im\left[ \frac{\lambda_b}{\lambda_s} C_7\right]
  \left[\cos \delta_\phi - \frac{q^2-m_\phi^2}{m_\phi \Gamma_\phi} \sin \delta_\phi
  \right] \,.\nn
\end{align}
The imaginary part in the above expression can be approximated as
$\Im[\lambda_b C_7]/\Re \,\lambda_s$.  When considering numerics in what
follows we will set $\Im [\lambda_b C_7]$ to the benchmark value of
$0.8\E{-2}$ in order to illustrate largest possible CP
effect. Relative importance of the $\cos \delta_\phi$ and $\sin
\delta_\phi$ for representative choices of $\delta_\phi$ is shown on
the upper plot in fig.~\ref{fig:aCPdiff}. The linearly rising
behaviour of the $\sin \delta_\phi$-driven term of the asymmetry is
compensated by a rapid drop of the resonant amplitude~\eqref{eq:BW}
that severely diminishes number of experimental events as we move
several $\Gamma_\phi$ away from $m_{\ell\ell} = m_\phi$. Both effects
are included in the effective experimental sensitivity that also takes
into account the rate of events in the considered kinematical region
and is shown on the bottom plot of fig.~\ref{fig:aCPdiff}. There we
plot $a_\mrm{CP}(m_{\ell\ell})$, weighted by the differential
branching ratio, a combined quantity that scales as $\sim
\mc{A}^\phi_\mrm{LD}\,\Im \mc{A}^\mrm{CPV}_\mrm{SD}$. These
sensitivity curves expose entirely different behaviour than
$a_\mrm{CP}(m_{\ell\ell})$. If the phase $\delta_\phi$ is close to $0$
or $\pi$ one finds the best sensitivity close to the peak. On the
contrary, for $\delta_\phi \sim \pm \pi/2$, the CP asymmetry is an
odd-function with respect to the resonant peak position and is maximal
when we are slightly off the peak. Therefore, experiment collecting
events in a symmetric bin around $m_{\ell \ell} = m_\phi$ would be
unable to observe CP asymmetry for maximal phase $\delta \sim \pm
\pi/2$.

\subsection{Partial-width CP asymmetries}
In order to keep the experimental search as general as possible one
should use appropriate search strategies to address the two limiting
possibilities, i.e. $\delta_\phi = 0,\pi$ and $\delta_\phi = \pm
\pi/2$. First, let us define a CP asymmetry of a partial width in the range
$m_1< m_{\ell\ell} < m_1$,
\begin{equation}
\label{eq:Acp}
  A_\mrm{CP} (m_1,m_2) = 
  \frac{\Gamma(m_1< m_{\ell\ell} < m_2)-\bar \Gamma(m_1 < m_{\ell\ell} < m_2)}{\Gamma(m_1< m_{\ell\ell} < m_2)+\bar \Gamma(m_1 < m_{\ell\ell} < m_2)}\,,
\end{equation}
where $\Gamma$ and $\bar \Gamma$ denote partial decay widths of $D^+$
and $D^-$ decays, respectively, to $\pi^\pm \mu^+ \mu^-$.  $A_\mrm{CP}$
is related to the differential asymmetry $a_\mrm{CP}(\sqrt{q^2})$
as
\begin{equation}
  A_\mrm{CP}  (m_1,m_2) =
  \frac{\int_{m_1^2}^{m_2^2} dq^2 R(q^2) \,a_\mrm{CP} (\sqrt{q^2})}{\int_{q_\mrm{min}^2}^{q_\mrm{max}^2} dq^2 R(q^2)}\,,
\end{equation}
where
\begin{equation}
  \label{eq:13}
  R(q^2) = \frac{1}{(q^2-m_\phi^2)^2+m_\phi^2 \Gamma_\phi^2}
  \int_{s_\mrm{min}(q^2)}^{s_\mrm{max}(q^2)} ds\, \sum_{s_+,s_-} \left|\bar u^{(s_-)}(k_-)
  \,\slashed{p} \,v^{(s_+)}(k_+)\right|^2
\end{equation}
involves the resonant shape and the
integral of the lepton trace over the Dalitz variable $s\equiv (p^\prime +
k_-)^2$ whose kinematical limits read
\begin{align}
  s_\mrm{max/min}(q^2) &= \frac{(m_D^2-m_\pi^2)^2}{4q^2} - \frac{\left(q^2
      \sqrt{1-\frac{4m_\mu^2}{q^2}} \mp
      \lambda^{1/2}(q^2,m_D^2,m_\pi^2)\right)^2}{4q^2}\,,\\
\lambda(x,y,z) &= (x+y+z)^2-4(xy+yz+zx)\,.\nn
\end{align}

\begin{figure}[!!hct]
  \centering
  \begin{tabular}{cc}
  \includegraphics[width=0.45\textwidth]{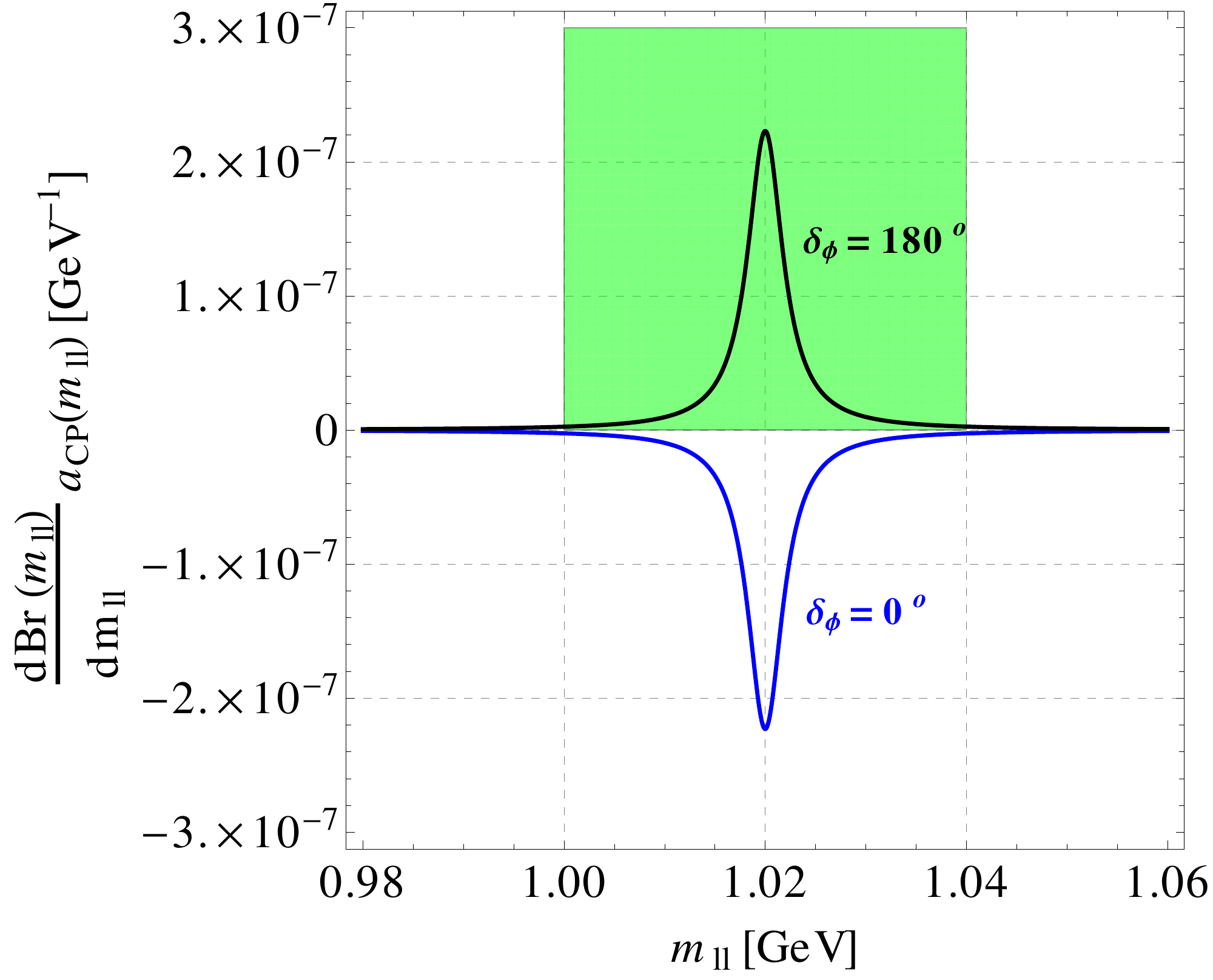} & \includegraphics[width=0.45\textwidth]{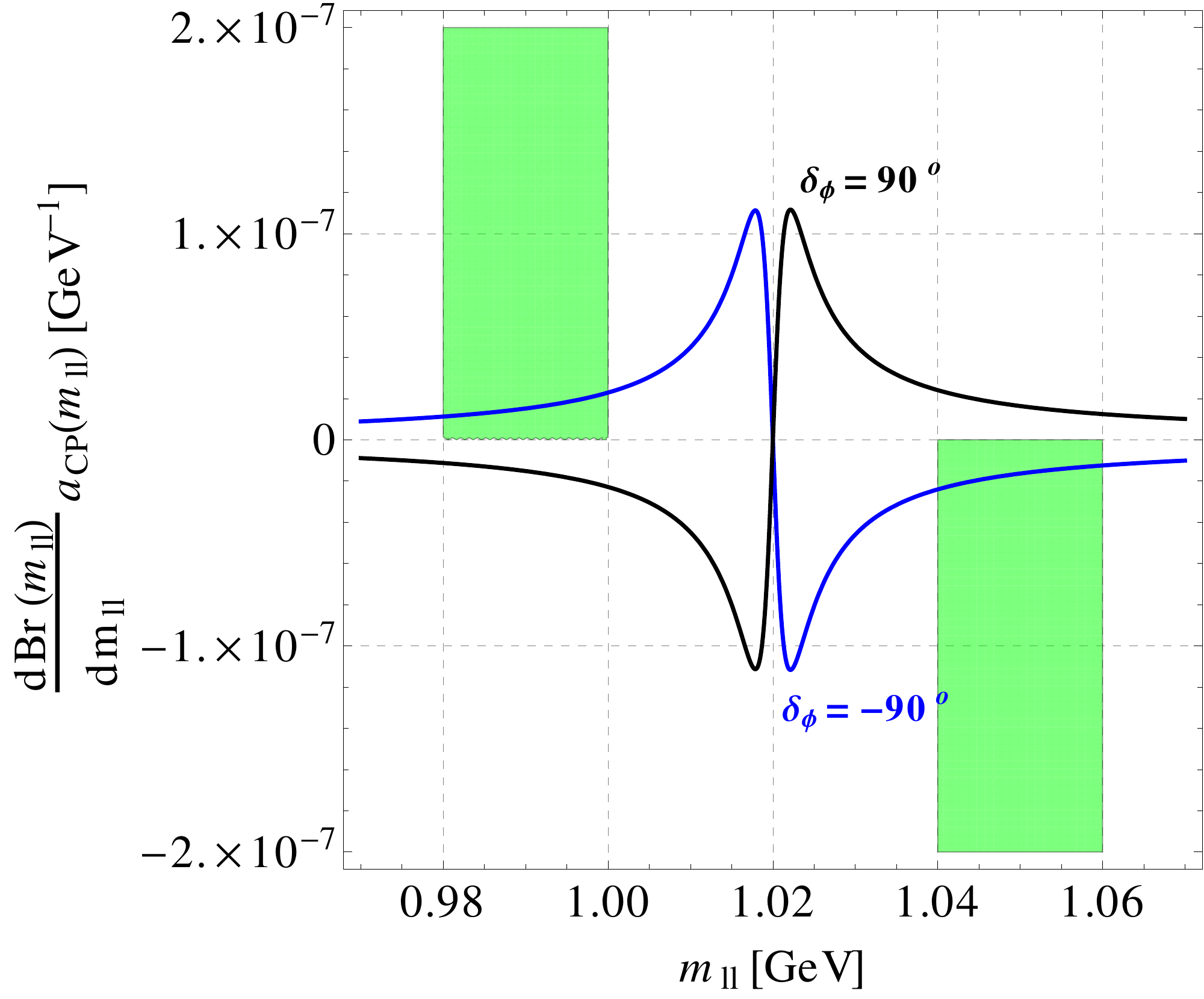}
  \end{tabular}
  \caption{Left: Asymmetry $a_{CP}(m_{\ell\ell})$ weighted by
    $d\mrm{Br}/dm_{\ell\ell}$ in the case when dominated by
    $\cos\delta_\phi$ term. The shaded region denotes the defining bin
    for asymmetry $C_{CP}^\phi$. Right: $a_{CP}(m_{\ell\ell})$ when
  dominated by $\sin \delta_\phi$. Shown are also the two bins where
  the asymmetry $S_{CP}^\phi$ is defined as the difference of $A_{CP}$
  in the two bins.}
  \label{fig:CSbins}
\end{figure}

The $D^+ \to \pi^+ e^+ e^-$ decay mode been searched for by the CLEO
experiment~\cite{Rubin:2010cq} where signal in a bin around the
$\phi$ resonance was observed. The following partial branching
ratio was reported
\begin{equation}
\mrm{Br}(D^+ \to \pi^+ e^+ e^-)_{|m_{ee}-m_{\phi}| \leq 20\e{MeV}} =
(1.7\pm 1.4\pm 0.1)\E{-6}\,,
\end{equation}
in a bin up covering the region $\sim 5\, \Gamma_\phi$ to the left and
right from the nominal position of the $\phi$ resonance. We define the
asymmetry on same bin for the $\pi^+ \mu^+ \mu^-$ final state as
\begin{equation}
\label{eq:S}
  C^{\phi}_\mrm{CP} \equiv A_{CP} (m_\phi - 20 \e{MeV}, m_\phi + 20 \e{MeV})\,.
\end{equation}
The asymmetry $C^\phi_\mrm{CP}$ is most sensitive to the $\cos
\delta_\phi$ term in Eq.~\eqref{eq:acp} and is therefore optimized for
cases when $\delta_\phi \sim 0$ or $\delta_\phi \sim \pi$.  Its
sensitivity would decrease if we approached $\delta_\phi \sim \pm\pi/2$, since the
$a_{CP}(m_{\ell\ell})$ would be asymmetric in $(m_{\ell\ell}-m_\phi)$ in
this case. For that
very region of $\delta_\phi$ we find the following observable with good
sensitivity to direct CP violation
\begin{align}
  \label{eq:A}
    S^{\phi}_\mrm{CP} &\equiv  A_{CP}(m_\phi - 40 \e{MeV}, m_\phi -
    20 \e{MeV}) -  A_{CP}(m_\phi + 20 \e{MeV}, m_\phi + 40 \e{MeV})
\end{align}
The bins where the partial width CP asymmetries $C^{\phi}_\mrm{CP}$
and $S^{\phi}_\mrm{CP}$ are defined are shown in fig.~\ref{fig:CSbins}
together with $a_{CP}(m_{\ell\ell})$.

\subsection{Case study for $C_\mrm{CP}^\phi$ and $S_\mrm{CP}^\phi$}
\begin{figure}[!!hct]
  \centering
  \begin{tabular}{cc}
      \includegraphics[width=0.45\textwidth]{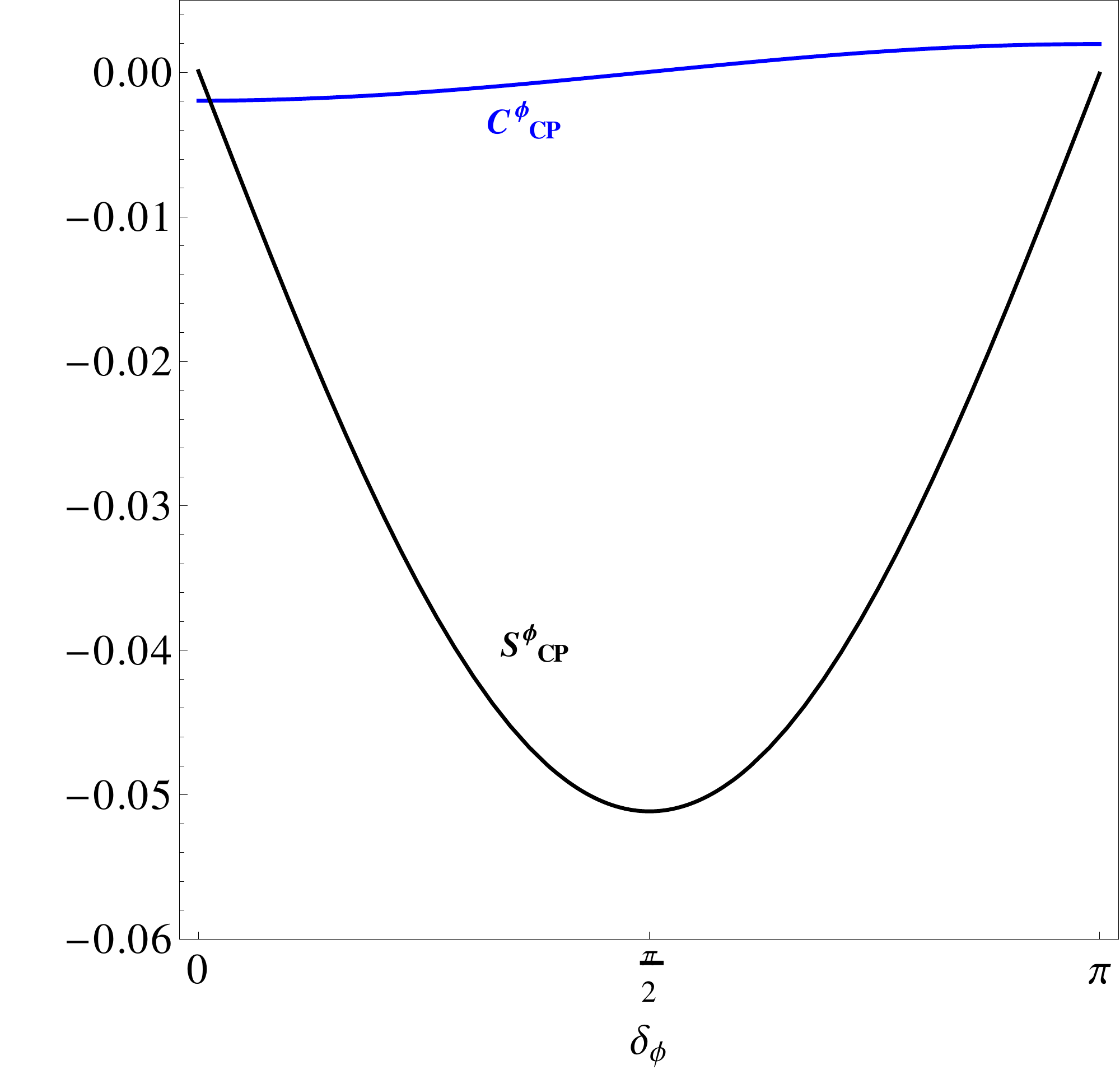} &       \includegraphics[width=0.49\textwidth]{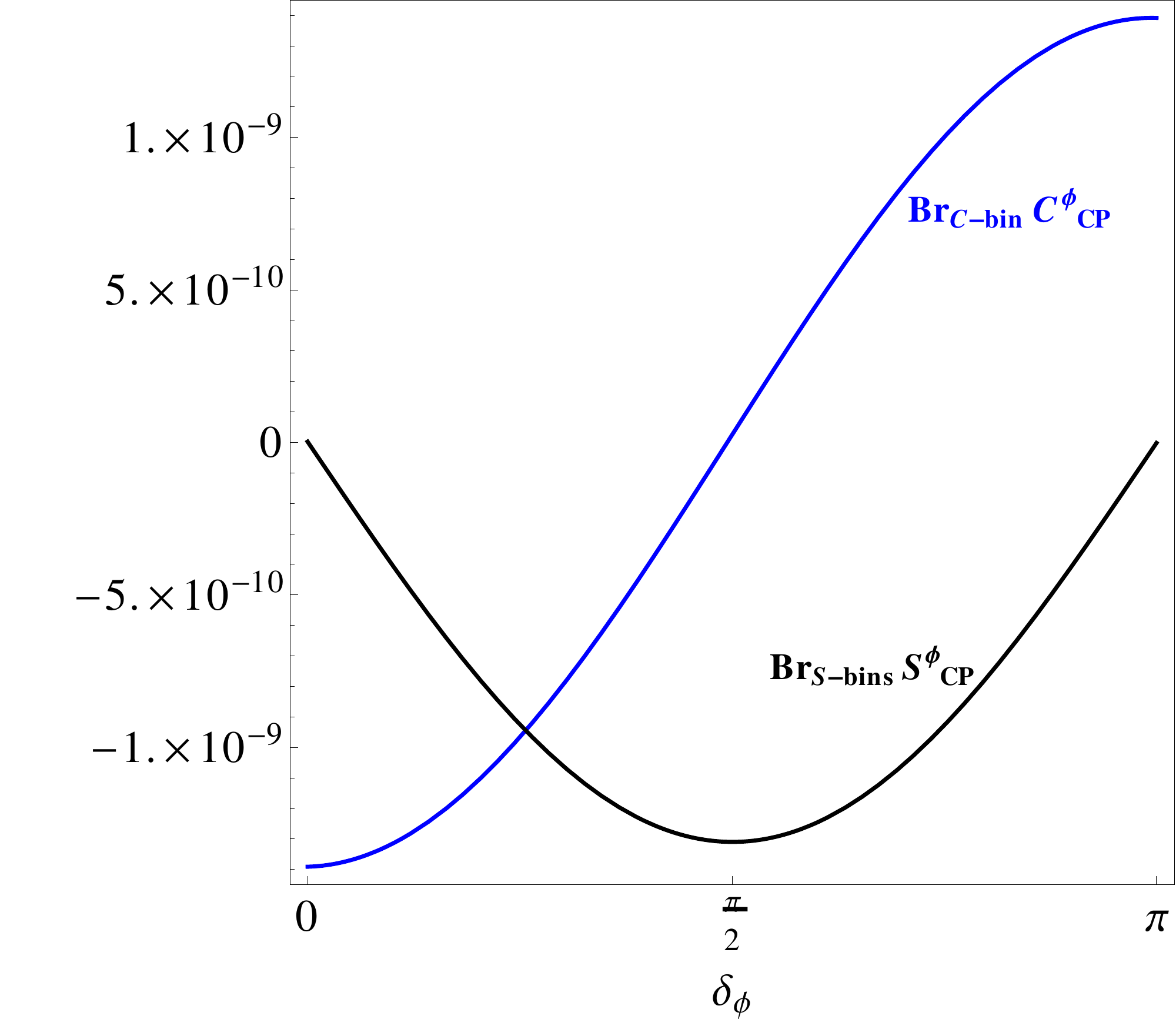}
  \end{tabular}
  \caption{Partial width asymmetries of $D \to \pi^+ \ell^+ \ell^-$
    decay. Left: asymmetries $C_{CP}^\phi$ and $S_{CP}^\phi$ for
    $\Im C_7 = 0.8\E{-2}$ and their dependence on $\delta_\phi$. Right:
    asymmetries rescaled by the branching ratios in the corresponding
    bins, thus representing effective sensitivity to direct CP violation.}
  \label{fig:CS}
\end{figure}

The asymmetry $S_\mrm{CP}^\phi$ can be an order of magnitude bigger
than $C_\mrm{CP}^\phi$ (see fig.~\ref{fig:CS}, left). However, when
we rescale the asymmetries by the branching ratios in the bins where
these asymmetries defined, namely by $7.1\E{-7}$ for $C_\mrm{CP}^\phi$
and $6.7\E{-8}$ for $S_\mrm{CP}^\phi$, we find evenly distributed
sensitivity to direct CP violation over entire range of
$\delta_\phi$. Also in the transient regions between the regimes where
either $\cos \delta_\phi$ or $\sin\delta_\phi$ terms dominate the
sensitivity does not decrease significantly. Numerical values of the
central values are summarized in tab.~\ref{tab:results}, whereas the
errors coming dominantly from parameter $a_\phi$~\eqref{eq:aphi} and
the form factor $f_T$~\eqref{eq:fTtildeBK} are estimated
to be of the order $20\,\%$.
\begin{table}[!hctbp]
  \centering
  \begin{tabular}{||c||c|c||c|c||}\hline\hline
   $\delta_\phi$ & $C_\mrm{CP}^\phi \times 10^2$ & $S_\mrm{CP}^\phi
   \times 10^2$ & $\mrm{Br}(\trm{C-bin}) \,C_\mrm{CP}^\phi \times 10^7$
   & $\mrm{Br}(\trm{S-bin}) \,S_\mrm{CP}^\phi \times 10^7$
   \\\hline
    $0$,$\pi$ & $\mp 0.20$ & $\pm 0.008$ & $\mp 0.014$ & $\pm 2\E{-5}$\\
    $\pm \pi/2$ & $\pm 0.003$ & $\mp 5.1$ & $\pm 2.4\E{-4}$ & $\mp 0.013$\\\hline\hline
  \end{tabular}
  \caption{Values of $D \to \pi^+ \mu^+ \mu^-$ CP asymmetries $C_\mrm{CP}^\phi$ and $S_\mrm{CP}^\phi$ for representative values of $\delta_\phi$. Last two columns show effective sensitivity.}
  \label{tab:results}
\end{table}

\subsection{Comment on $D_s \to \phi K^+ \to K^+ \ell^+ \ell^-$}
Same type of asymmetries can be defined for the decay mode of $D_s$
meson via the $\phi$ resonance to final state $K^+ \ell^+ \ell^-$.
The resonant amplitude is described by an analogous expression
to~\eqref{eq:BW} and is parameterized by real $a_\phi^\prime$ and
$\delta_\phi^\prime$. The branching ratio
\begin{align}
 \mrm{Br}(D_s^+ \to \phi K^+) &= (1.8\pm 0.4 )\times 10^{-4}\,,
\end{align}
obtained from $\mrm{Br}(D_s^+ \to \phi(\to K^+ K^-) K^+) = (9.0 \pm
2.1)\E{-5}$ and $\mrm{Br}(\phi \to K^+ K^-) = 0.489 \pm
0.005$~\cite{Nakamura:2010zzi}, is an order of magnitude smaller than
the corresponding $\mrm{Br}(D^+ \to \phi \pi^+)$. By employing the
narrow width approximation the value we find $a_\phi^\prime = 0.49$
with $\sim 10\,\%$ error. On the other hand, the short distance
amplitude remains of same order of magnitude as in the $D^+ \to \pi^+
\mu^+ \mu^-$ case. We neglect the $SU(3)$-breaking corrections to the
form factor and use $f_T(q^2)$ as given in~\eqref{eq:fTtildeBK} adjusted by
$m_\pi \to m_K$.
\begin{figure}[!!hct]
  \centering
  \begin{tabular}{cc}
      \includegraphics[width=0.45\textwidth]{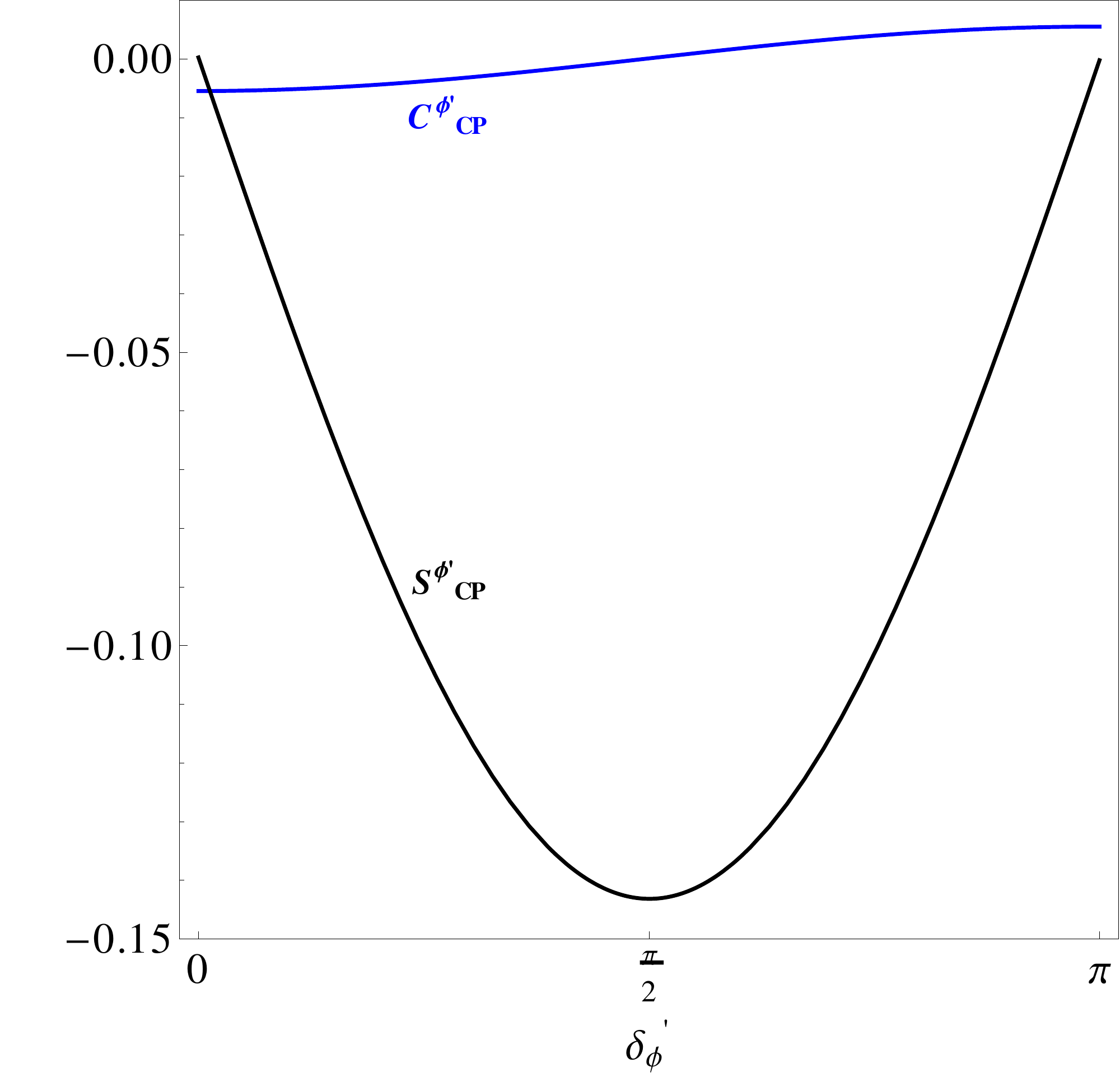} &       \includegraphics[width=0.49\textwidth]{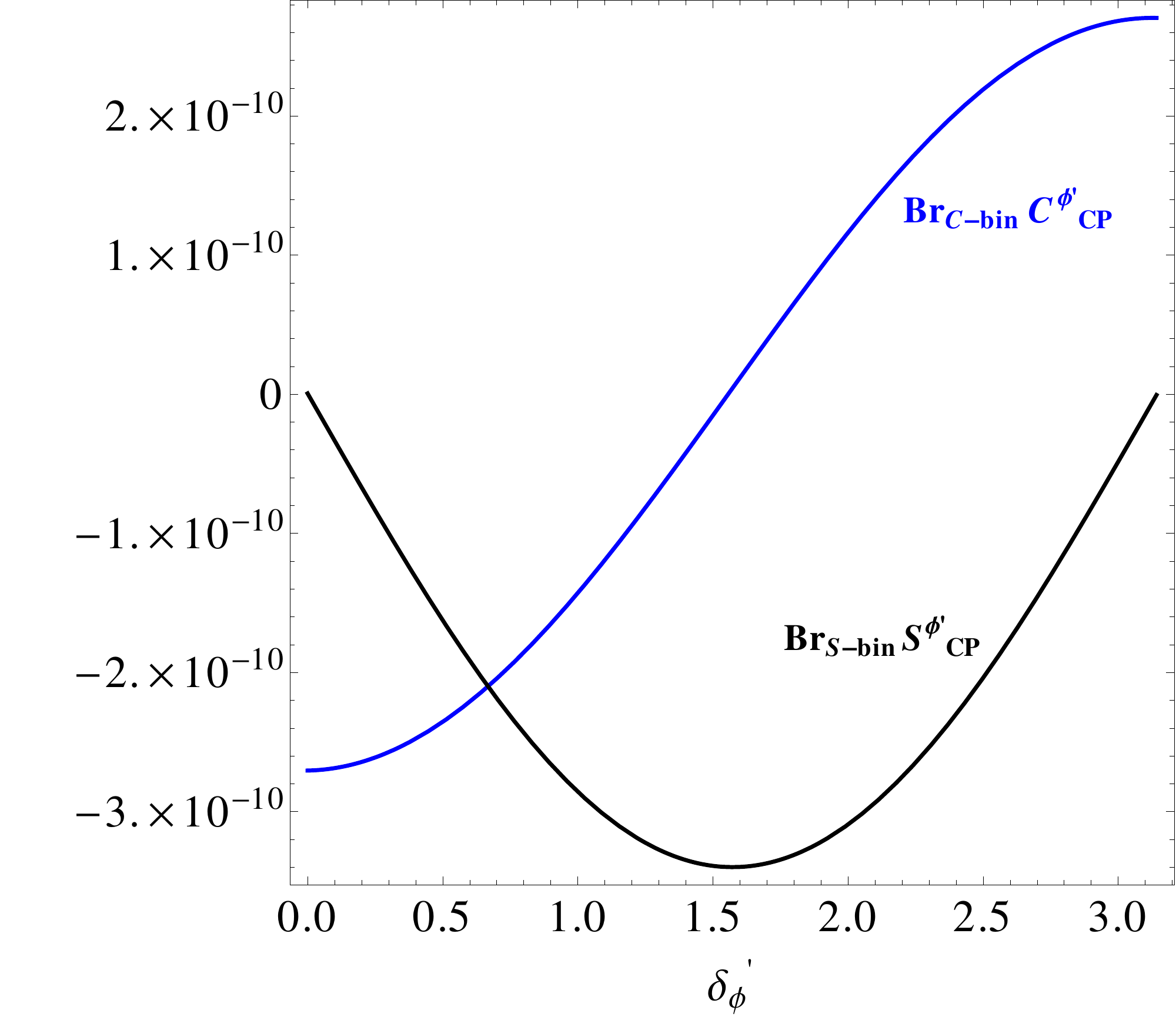}
  \end{tabular}
  \caption{Partial width asymmetries of $D_s \to K^+ \ell^+ \ell^-$
    decay. Left: asymmetries $C_{CP}^{\phi\prime}$ and $S_{CP}^{\phi\prime}$ for $\Im
    C_7 = 0.8\E{-2}$ and their dependence on $\delta_\phi^\prime$. Right: asymmetries
    rescaled by the branching ratios in the corresponding bins, thus
    representing effective sensitivity to direct CP violation.}
  \label{fig:CS-s}
\end{figure}
The asymmetries $C_\mrm{CP}^{\phi\prime}$ and
$S_\mrm{CP}^{\phi\prime}$ are larger, whereas the experimental
sensitivity is weaker due to smaller branching fractions, as shown in
tab.~\ref{tab:resultsS}.
\begin{table}[!hctbp]
  \centering
  \begin{tabular}{||c||c|c||c|c||}\hline\hline
   $\delta_\phi^\prime$ & $C_\mrm{CP}^{\phi\prime} \times 10^2$ & $S_\mrm{CP}^{\phi\prime}
   \times 10^2$ & $\mrm{Br}(\trm{C-bin}) \,C_\mrm{CP}^{\phi\prime} \times 10^7$
   & $\mrm{Br}(\trm{S-bin}) \,S_\mrm{CP}^{\phi\prime} \times 10^7$
   \\\hline
    $0$,$\pi$ & $\mp 0.55$ & $\pm 0.024$ & $\mp 0.0027$ & $\pm 1\E{-5}$\\
    $\pm \pi/2$ & $\pm 0.008$ & $\mp 14$ & $\pm 4\E{-5}$ & $\mp 0.007$\\\hline\hline
  \end{tabular}
  \caption{Values of $D_s \to K^+ \mu^+ \mu^-$ CP asymmetries
    $C_\mrm{CP}^{\phi\prime}$ and $S_\mrm{CP}^{\phi\prime}$ for representative values of $\delta_\phi^\prime$. Last two columns show effective sensitivity.}
  \label{tab:resultsS}
\end{table}

\section{Summary}
\label{sec-summ}
In this article we have studied CP asymmetries of rare decays $D^+ \to
\pi^+ \mu^+ \mu^-$ and $D_s \to K^+ \mu^+ \mu^-$ defined close to the
$\phi$ resonance that couples to the lepton pair. These asymmetries
can be generated by imaginary parts of Wilson coefficients in the
effective Hamiltonian for $c \to u \ell^+ \ell^-$ processes. We have
limited the discussion to the electromagnetic dipole coefficient $C_7$
which can carry a large CP-odd imaginary part, if the direct CP
violation in singly Cabibbo suppressed decays $D \to \pi \pi, KK$ is to be
explained by NP contribution to the chromomagnetic operator $\mc{O}_8$.

We have focused on the CP asymmetry around the $\phi$ resonant peak in
spectrum of dilepton invariant mass. There approximate description of
the resonant amplitude by means of the Breit-Wigner ansatz with two
additional parameters is expected to dominate over all other CP
conserving contributions. Possible long-distance CP violating
contributions of chromomagnetic operator have been neglected in this
work. We have fixed one of the resonance parameters from the known
resonant branching fractions of $D_{(s)} \to \phi (\to \mu^+ \mu^-)
P$, while the remaining parameter is an unknown CP-even strong phase
$\delta_\phi$. The resonant amplitude in addition generates a phase
that depends on the dilepton invariant mass. The hadronic dynamics of
the short distance part of the amplitude is contained in a tensor form
factor, $f_T$, that has been calculated in quenched lattice simulations of QCD.

The interference term between the resonant and the short distance
amplitude that drives the direct CP asymmetry depends decisively on
the particular value of the strong phase. Namely, for large strong
phase $\delta_\phi$, i.e., close to either $+\pi/2$ or $-\pi/2$, the
CP asymmetry would vanish should the experimental bin enclose the
$\phi$ peak symmetrically. Conversely, the same CP asymmetry would be
most sensitive when the strong phase was either close to $0$ or
$\pi$. In order to cover experimentally the whole range of strong
phase values we have devised two asymmetries that are maximally
sensitive either to peak-symmetric or peak-antisymmetric CP
violation. Taking $0.008$ for the imaginary part of $V_{cb}^* V_{ub} C_7$, the two
asymmetries can take values of the order $10\,\%$ for $\delta_\phi =
\pm\pi/2$ or of the order $0.1-1\,\%$ for $\delta_\phi = 0,\pi$. When
we multiply the asymmetries by the partial branching fractions in the
corresponding bins, the two asymmetries provide an almost even
sensitivity for all values of the strong phase. For the $D \to \pi^+
\mu^+ \mu^-$ thus defined sensitivity amounts to $\sim 1\E{-9}$ and
$\sim 3\E{-10}$ for $D_s \to K^+ \mu^+ \mu^-$, bearing in mind that CP
asymmetry and experimental sensitivity are proportional to the
imaginary part of $C_7$. We conclude that measurements of partial
width CP asymmetries in decays $D^+ \to \pi^+ \mu^+ \mu^-$ and $D_s^+
\to K^+ \mu^+ \mu^-$ might be useful in investigating whether new
physics in chromomagnetic operator is responsible for direct CP
violation in singly Cabibbo suppressed decays to two pseudoscalar
mesons.

\begin{acknowledgments}
  N.~K. thanks B.~Viaud for discussions about the experimental
  sensitivity in charm semileptonic decays. We thank D.~Be\v cirevi\'c
  for providing lattice results for the tensor form factor and sharing
  with us useful comments regarding form factors and long distance
  amplitude. This work is supported in part by the Slovenian Research
  Agency. N.~K. acknowledges support by {\sl Agence Nationale de la
    Recherche}, contract LFV-CPV-LHC ANR-NT09-508531.
\end{acknowledgments}

\bibliography{refs}
\end{document}